# Size Effect and Scaling in Quasi-static and Fatigue Fracture of Graphene Polymer Nanocomposites

*Yao Qiao[a], Kaiwen Guo[b], Marco Salviato[a,∗]*

[a]*William E. Boeing Department of Aeronautics and Astronautics, University of Washington, Seattle, Washington, 98195, USA*
[b]*School of Aerospace Engineering, Tsinghua University, Beijing, 100084, China*

**Abstract**

This work investigated how the structure size affects the quasi-static and fatigue behaviors of graphene polymer nanocomposites, a topic that has been often overlooked. The results showed that both quasi-static and fatigue failure of these materials scale nonlinearly with the structure size due to the presence of a significant Fracture Process Zone (FPZ) ahead of the crack tip induced by graphene nanomodification. Such a complicated size effect and scaling in either quasi-static or fatigue scenario cannot be described by the Linear Elastic Fracture Mechanics (LEFM), but can be well captured by the Size Effect Law (SEL) which considers the FPZ.

Thanks to the SEL, the enhanced quasi-static and fatigue fracture properties were properly characterized and shown to be independent of the structure size. In addition, the differences on the morphological and mechanical behaviors between quasi-static fracture and fatigue fracture were also identified and clarified in this work.

The experimental data and analytical analyses reported in this paper are important to deeply understand the mechanics of polymer-based nanocomposite materials and even other quasi-brittle materials (*e.g.*, fiber-reinforced polymers or its hybrid with nanoparticles, etc.), and further advance the development of computational models capable of capturing size-dependent fracture of materials in various loading conditions.

1. Introduction

Graphene polymer nanocomposites have been widely studied owing to graphene's exceptional material properties [1-4]. In fact, a well-distributed state of graphene nanoparticles in a polymer matrix can significantly alter the movement paths of crack propagation and diffusion, and the transport behavior of electron and phonon.

---

∗Corresponding Author: salviato@aa.washington.edu

Accordingly, graphene polymer nanocomposites can exhibit outstanding specific physical properties, thus enabling several applications (*e.g.*, sensing [5], photovoltaics [6], gas barriers [7], electronics [8], fiber-reinforced polymers with graphene nanomodification [9-11], graphene-reinforced adhesives for materials joining [12,13], etc.).

Following the main focus of this paper on fracture properties, graphene nanoparticles with different forms (*e.g.*, nanoplatelets, nanosheets, nanotubes, nanoribbons, graphene oxide, functionalized graphene, etc.) embedded in polymer matrices have been often shown to exhibit enhanced fracture resistance compared to that of pure polymer matrices. The main enhancement mechanisms (*e.g.*, [14-18], etc.) typically can be summarized into the following types: (1) crack deflection; (2) crack pinning and bifurcation; (3) graphene pull-out; (4) crack bridging; (5) separation of graphene layers. Some insightful studies for fracture of graphene polymer nanocomposites were described next.

Among others [19-23], an interesting work was reported by Shokrieh *et al.* [24] who showed that a 20-40% increase of the Mode I fracture toughness can be achieved with only 0.5wt% graphene nanoplatelets or nanosheets embedded into an epoxy matrix. A higher enhancement of the Mode I fracture energy more than 90% compared to that of pure epoxy was achieved by using 1.6wt% graphene nanoplatelets as reported by Mefford *et al.* [25]. By using other forms of graphene [26-31], for instance, Shirodkar *et al.* [32] showed that an epoxy with the reinforcement of only 0.5wt% graphene nanotubes can have the Mode I fracture toughness being at least two times higher than that of pure epoxy. It was also reported by Shirodkar *et al.* [32] that a hybrid modification of a polymer matrix with the combination of both graphene nanoplatelets and nanotubes can improve the Mode I fracture toughness more than 100% compared to that of pure epoxy. In addition to thermoset-based nanocomposites, graphene nanomodification was also shown to improve the Mode I fracture toughness of different thermoplastic polymers [33-40]. For instance, Cicero *et al.* [41] investigated the Mode I fracture behavior of additively manufactured graphene-nanoplatelets-reinforced polylactic acid. It was found that the enhanced fracture resistance due to the addition of 1wt% graphene nanoplatelets depends on both raster orientation and notch radius, and the highest improvement can be more than 50% compared to pure polylactic acid with same configuration. For other fracture modes (*e.g.*, Mode II, mixed-mode, etc.), an outstanding study was conducted by Kumar and Roy [42,43] who investigated both pure-mode and mixed-mode fracture of graphene polymer nanocomposites via asymmetrical four-point bending tests. Their results showed that the addition of 0.5wt% graphene nanoplatelets into an epoxy can lead to more than 150% improvement on both pure Mode II and mixed-mode fracture roughness compared to the baseline, and the enhanced mixed-mode fracture envelope for different graphene contents were also constructed. Few studies on Mode II and mixed-mode fracture of graphene-based nanocomposites can also be found in the literature [44,45].

Regarding the fatigue (cyclic) behavior of graphene polymer nanocomposites, different forms of graphene have also been studied in the literature to explore their effects on the potential improvements of fatigue crack initiation and propagation of polymer matrices (*e.g.*, [46-50], etc.). Heieh *et al*. [51] added 0.5wt% graphene nanotubes into an epoxy and found an increase of the Mode I fatigue threshold about 30% compared to that of pure epoxy. However, the evolution of the crack growth rate of a graphene-nanomodified epoxy in fatigue remains almost the same as that of pure epoxy. Interestingly, Rafiee *et al*. [52] reported a noticeable reduction of the Mode I fatigue crack growth rate of an epoxy due to the addition of 0.125wt% graphene nanosheets. Moreover, the effect of graphene nanoparticles on the stress-life behavior (*i.e.*, S-N or Wöhler's curve [53]) of polymer matrices was also studied in the literature [48,54]. For instance, Jen *et al.* [55] showed a shift of S-N curves of graphene polymer nanocomposites towards the direction of improving the fatigue lifetime to failure by embedding 0.4wt% nanotubes, nanoplatelets, or different ratios of both nanoparticles into an epoxy system. However, it was also shown by Jen *et al.* [55] that the investigated graphene nanoparticles do not change the slope of the S-N curve of a pure epoxy. It is worth mentioning here that the Mode II and mixed-mode fatigue (cyclic) fracture of graphene polymer nanocomposites were extremely lack of studies in the literature, which may need more investigations to better understand the fatigue failure behavior of these materials under complex loads.

Although the foregoing efforts led to a remarkable progress in understanding the damage and fracture of graphene polymer nanocomposites under either quasi-static or fatigue loading condition, the size effect and scaling of quasi-static and fatigue fracture in these materials were rarely explored in the literature due to the complexity of the problems. However, this research topic is very important for a broad application of graphene-based nanocomposites at different length scales and loading conditions. In particular, the size-dependent fatigue (cyclic) damage and fracture of graphene polymer nanocomposites were not investigated in the literature, and its difference compared to quasi-static fracture was also not clear yet in the literature as it can be indicated from some studies on the fatigue scaling of other quasi-brittle materials (*e.g.*, concrete [56-59], rocks [60,61], etc.).

This work aims to fill the foregoing knowledge gap by investigating the size effect and scaling of graphene-nanoplatelets-reinforced thermosetting polymers both in quasi-static and fatigue loading conditions as an example. It was clearly shown that the damage and fracture of these nanocomposites strongly depend on the structure size. The size-dependent quasi-static and fatigue failure characteristics of graphene polymer nanocomposites were identified, explained, and also captured by the Size Effect Law (SEL) [62,63] with the proposed modification for fatigue fracture analysis. The results in this work can be used for the calibration, validation, and development of advanced computational models. Also, this study particularly provides a better understanding of size-dependent fatigue fracture of graphene polymer nanocomposites and other polymer-based quasi-brittle materials (*e.g.*, nanocomposites with other reinforcing fillers, fiber-polymer composites, fiber-particle-polymer hybrid composites, etc.), which was often overlooked and less studied in the literature.

## 2. Materials, Specimens, Testing, and Characterizations

Two material systems were investigated in this work: (a) pure epoxy resin and (b) epoxy resin toughened by the addition of graphene nanoplatelets. Three different graphene contents (0wt%, 0.9wt%, and 1.6wt%) were selected to study the effects of quasibrittleness induced by graphene nanomodification on both quasi-static and fatigue failure behavior. A detailed description of the materials and their preparation is provided next.

### 2.1. Materials

The thermoset polymer used in this study was composed of an EPIKOTE$^{TM}$ Resin MGS$^{TM}$ and an EPICURE$^{TM}$ Curing Agent MGS$^{TM}$ RIMH 137 combined in a 100:32 ratio by weight [64]. The investigated nanofiller was A-12 graphene nanoplatelet with lateral dimensions of about 2-8 $\mu$m and an average thickness for a graphene monolayer of less than 1 nm [65].

#### 2.1.1. *Preparation of pure epoxy*

The manufacturing process for the thermoset polymer started from manually mixing the epoxy resin with the hardener for about 10 minutes, followed by degassing the mixture for about 20 minutes in a vacuum trap by using a mobile vacuum system [66] in order to remove any air bubbles. The mixture was then poured into a silicone mold made of RTV silicone from TAP Plastics [67] to achieve consistent geometries of the specimens. The last step of the manufacturing process involved the curing of the specimens at room temperature for about 48 hours and post-curing in a convection oven for about 4 hours at 60°C.

### 2.1.2. *Preparation of graphene nanocomposites*

For the preparation of graphene nanocomposites, the resin was manually mixed with graphene nanoplatelets for about 10 minutes and then mixed by means of a high shear mixer with a 48mm impeller [68] at 1500rpm for about 20 minutes. Furthermore, sonication was performed to promote nanoplatelet exfoliation by using a Hielscher UP200S sonicator [69] with a 7mm sonotrode for about 20 minutes at 70% amplitude and a duty cycle of 0.5. Similar procedures were successfully adopted by several authors in the literature for the exfoliation of other nanoparticles in thermoset polymers [70-72]. After adding the hardener and degassing the mixture, the procedure followed the same steps outlined for pure epoxy.

### 2.2. Specimens and testing

To study the size effect and scaling of the fracture behavior of graphene nanocomposites, as illustrated in Figure 1, geometrically scaled Single Edge Notch Bending (SENB) specimens of three different sizes were prepared for each material configuration. The design of the SENB specimens was based on ASTMD5045-99 [73]. The in-plane dimensions ($D \times L$) with the scaling of 1:2:4 were 10×36 mm, 20×72 mm, and 40×144 mm, respectively where $D$ is the specimen width and $L$ is the span between two supports. The scaling did not involve the thickness ($t$), which was 12 mm for all the investigated sizes.

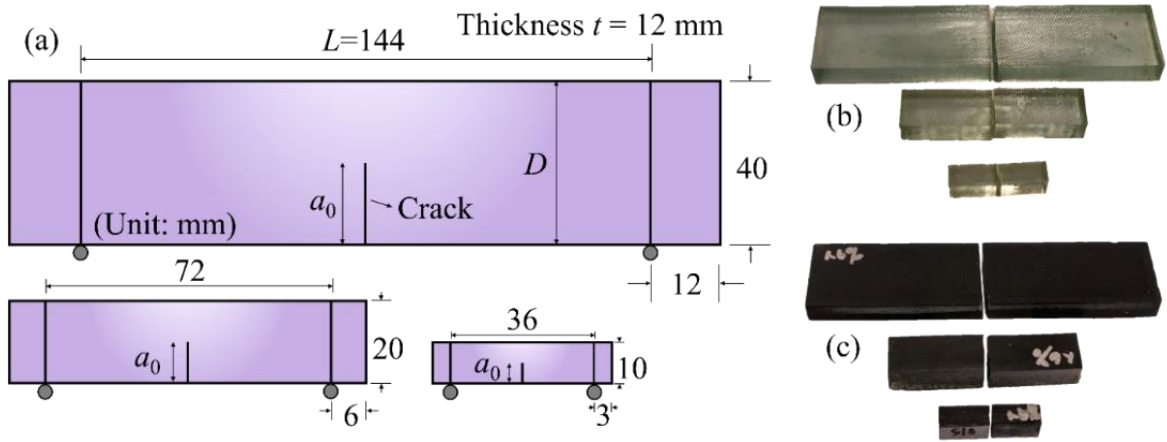

**Figure 1.** (a) Geometrically-scaled SENB specimens for Mode I tensile quasi-static and fatigue tests. Fractured specimens for (b) pure epoxy and (c) 1.6wt% graphene content as an example.

The initial, sharp cracks in the SENB specimens were made by tapping with a razor blade after sawing a pre-notch by means of a diamond-coated saw. Using tapping to create the last portion of the crack was extremely important to investigate the fracture properties of graphene nanocomposites since a blunt notch created by sawing can lead to a fracture only governed by the local material strength at the notch tip depending on the ratio of the notch radius to notch length as reported in the literature for various materials [74,75].

It is worth mentioning that achieving a consistent ratio of crack length to specimen width for all the specimen sizes and material configurations was difficult. Also, the specimens with graphene nanomodification were completely black opaque, even at the lowest concentration, it was difficult to identify the crack tip. Thus, the specimens were painted white so that the contrast in colors provided a better observation of the crack tip location after tapping. The initial crack length ($a_0$) of the SENB specimens was about 0.35-0.55$D$ for quasi-static fracture tests, whereas the initial crack length ($a_0$) was about 0.25-0.4$D$ for the SENB specimens used for fatigue fracture tests.

### 2.2.1. *Quasi-static fracture tests*

SENB specimen was placed on two metal supports with cylindrical pins of about 4 mm in diameter. The displacement on the top and middle of the specimen was applied by using a metal wedge with the cylindrical pin of the same size as used for supports. Quasi-static fracture tests were performed on a closed-loop electro-activated 5585H Instron machine under displacement control. The displacement rates were 0.05 mm/min, 0.08 mm/min, and 0.12 mm/min for the specimens with $D$ = 10 mm, 20 mm, and 40 mm respectively. These values were adjusted for the different specimen sizes to roughly achieve the same nominal strain rate ($\dot{\varepsilon}_n$), the nominal strain being defined as $\varepsilon_n = \sigma_n/E = 3PL/(2EtD^2)$ where $\sigma_n$ is the nominal stress, $P$ is the load, $E$ is the elastic modulus, and others have the same meaning as discussed in previous sections. This choice was made to mitigate possible strain-rate effects, which are known to affect the mechanical behavior of thermoset polymers.

### 2.2.2. *Fatigue (cyclic) fracture tests*

Same testing configuration as quasi-static fracture tests was adopted for fatigue fracture tests, which were conducted on a servo-hydraulic 8851 Instron machine under the load control. The same stress ratio $R = \sigma_{n,min}/\sigma_{n,max} = 0.1$ but with different loading frequencies were used for different specimen sizes to roughly have the same nominal strain rate for the mitigation of the strain rate effect as previously mentioned for quasi-static fracture tests. In the expression of the stress ratio, $\sigma_{n,min}$ and $\sigma_{n,max}$ represent minimum and maximum values of the nominal stresses as defined in Section 2.2.1. The loading frequencies were 4 Hz, 7.4 Hz, and 11Hz for the specimens with $D = 10$ mm, 20 mm, and 40 mm respectively. A digital microscope [76] with a frame rate of 30 fps was used to record the entire fatigue lifetime of the specimen for further analysis.

*2.3. Morphology characterizations*

To investigate the fracture morphologies and toughening mechanisms of graphene nanocomposites at the microscale, Scanning Electron Microscopy (SEM) was performed by means of a JSM-6010PLUS/LA Electron Microscope [77] under a voltage from about 10 to 15 kV. Before observation, the fracture surfaces taken from the geometrically scaled specimens were coated with a SC7620 Mini Sputter Coater/Glow Discharge System from Quorum Technologies [78].

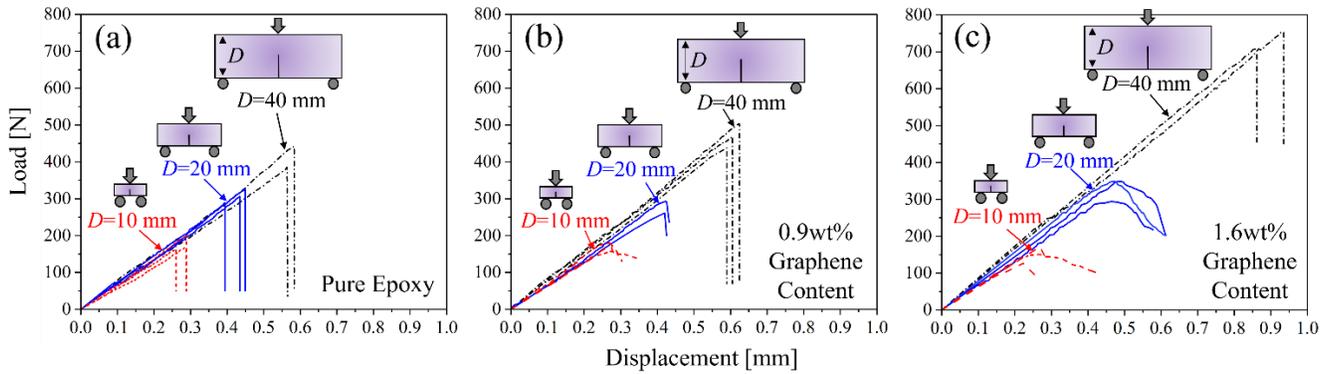

**Figure 2.** Load-displacement curves of the SENB specimens with different graphene contents obtained from the Mode I quasi-static fracture tests: (a) pure epoxy (0wt% graphene content); (b) 0.9wt% graphene content; (c) 1.6wt% graphene content. For each graphene content, three different geometrically scaled SENB specimens were tested in this work.

3. **Experimental Results and Discussion**

*3.1. Mode I fracture under quasi-static loading*

The load-displacement curves of the SENB specimens with different graphene contents obtained from the quasi-static fracture tests were plotted in Figure 2. For pure thermoset polymer, the mechanical behavior is linear up to the peak load and followed by rapid crack propagation. This indicates the brittleness of all the investigated sizes. By adding graphene nanoplatelets (Figure 2b-c), while large specimens still exhibit linearity up to failure, a significant non-linear portion before the peak load describes the smaller specimens. This phenomenon becomes much more pronounced as graphene content increases (Figure 2c), and indicates higher fracture ductility for higher graphene contents and smaller specimen sizes. After reaching the peak load, for all the investigated sizes and graphene contents, crack propagated unstably and led to a catastrophic failure of the specimen due to snap-back instability as often reported in the literature for various quasi-brittle materials [79,80].

*3.2. Mode I fracture under fatigue (cyclic) loading*

For different graphene contents and specimen sizes, the evolution of the normalized equivalent crack length ($a_{eq}/D$) with fatigue lifetime was plotted in a semi-logarithmic coordinate as illustrated in Figure 3. Equivalent crack length $a_{eq}$ was obtained by means of the compliance method. Details about how to obtain the equivalent crack length based on the specimen compliance can be found in Appendix 1.

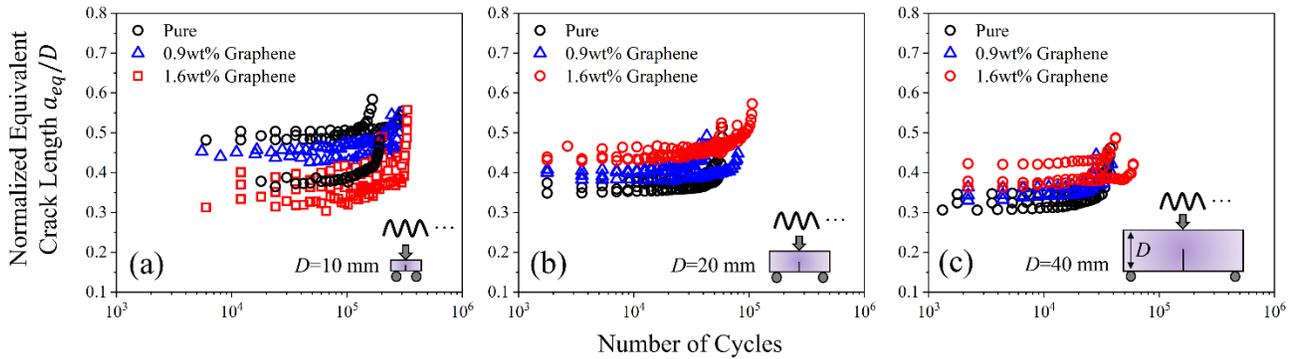

**Figure 3.** Normalized equivalent crack length $\alpha_{eq} = a_{eq}/D$ with respect to number of cycles obtained from the Mode I cyclic fracture tests on the SENB specimens with different sizes: (a) $D = 10$ mm; (b) $D = 20$ mm; (c) $D = 40$ mm. Each figure compares three different graphene contents (0wt%, 0.9wt%, and 1.6wt%). Note, details on the calculation of the equivalent crack length $a_{eq}$ can be found the Appendix 1.

As can be noted from Figure 3 for different graphene contents, sub-critical crack exhibited slow growth at the early stage of fatigue lifetime, whereas a rapid crack propagation dominated the fatigue fracture behavior close to the final failure of the specimen. Notwithstanding these similarities, the addition of graphene nanoplatelets led to the improvement of fatigue lifetime as shown in Figure 4a. It is worth mentioning here that the foregoing improvement is based on the fact that the applied cyclic loads on the SENB specimens with different graphene contents and sizes were very close to their fatigue thresholds ($\Delta K_{eq,th}$). By taking an example of 1.6wt% graphene nanoplatelets, fatigue lifetime of the SENB specimen was improved about 28%, 39%, and 34% compared to the pure epoxy for the specimen with $D = 10$ mm, 20 mm, and 40 mm respectively. It is not surprising since the existence of graphene nanoplatelets can retard the crack propagation and cause a tortuous crack path as it will be shown and described in the next section. On the other hand, in

Figure 4b, fatigue lifetime increases in a non-linear way as the specimen size decreases for all the investigated graphene contents. This indicates an increased quasi-brittleness for a smaller specimen size as it will be shown in Section 4.2. The details on the explanation of this phenomenon must leverage a computational modeling to correlate cyclic stress evolution in front of the crack tip with different specimen sizes, which is beyond the scope of this study and considered as future publications.

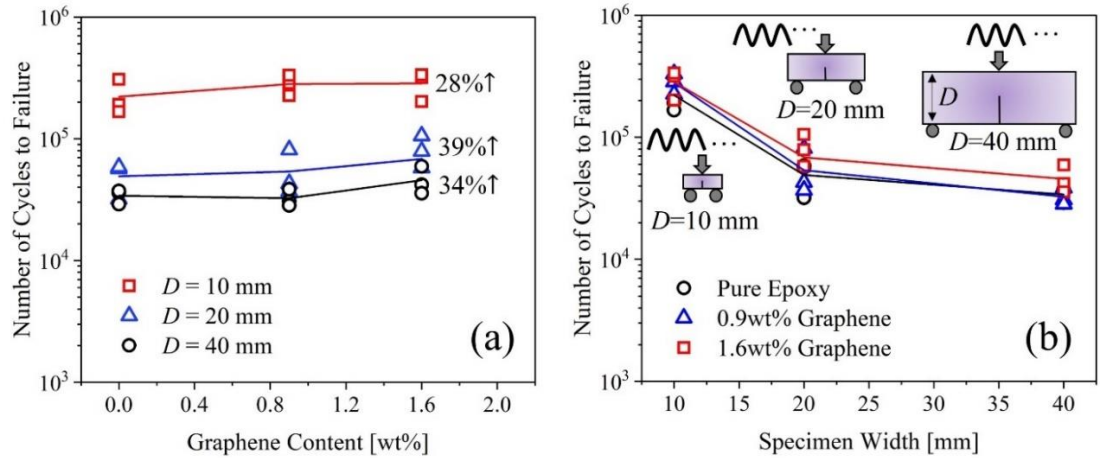

**Figure 4.** (a) Number of cycles to failure as a function of graphene content for different specimen sizes; (b) number of cycles to failure as a function of specimen size for different graphene contents.

*3.3. Fracture morphologies and toughening mechanisms*

*3.3.1. Quasi-static condition*

Surface morphologies of the SENB specimens with different graphene contents after quasi-static failure were investigated as shown in Figure 5. As can be noted from Figure 5a-c highlighting the differences among different graphene contents, the fracture surface of the pure epoxy shows smoothness, whereas the addition of the graphene nanoplatelets leads to a rougher texture of the surface after fracture. Surface roughness increases as the graphene content increases as illustrated in Figure 5a-c for the comparison among pure epoxy, 0.9wt% graphene content, and 1.6wt% graphene content.

To understand the microscale mechanisms of damage in graphene polymer nanocomposites under quasi-static loading condition, higher magnification images in the crack propagation region of the SENB specimens with 0.9wt% and 1.6 % graphene contents were investigated and illustrated in Figure 5d-f. According to the micrographs, three different mechanisms can be identified: (a) crack deflection into a different plane after meeting graphene nanoplatelets (Figure 5d); (b) pull-out of graphene nanoplatelets due to crack opening (Figure 5e); (c) crack pinning and bifurcation after meeting graphene nanoplatelets (Figure 5f). In addition, crack can also propagate in between the graphene layers and then split the agglomerate into two parts [81]. Similar damage characteristics were also shown and reported in the literature for nanofiller-reinforced polymers [82,83]. All the forgoing damage mechanisms cause a more torturous damage path thus dissipating more energy during the crack propagation.

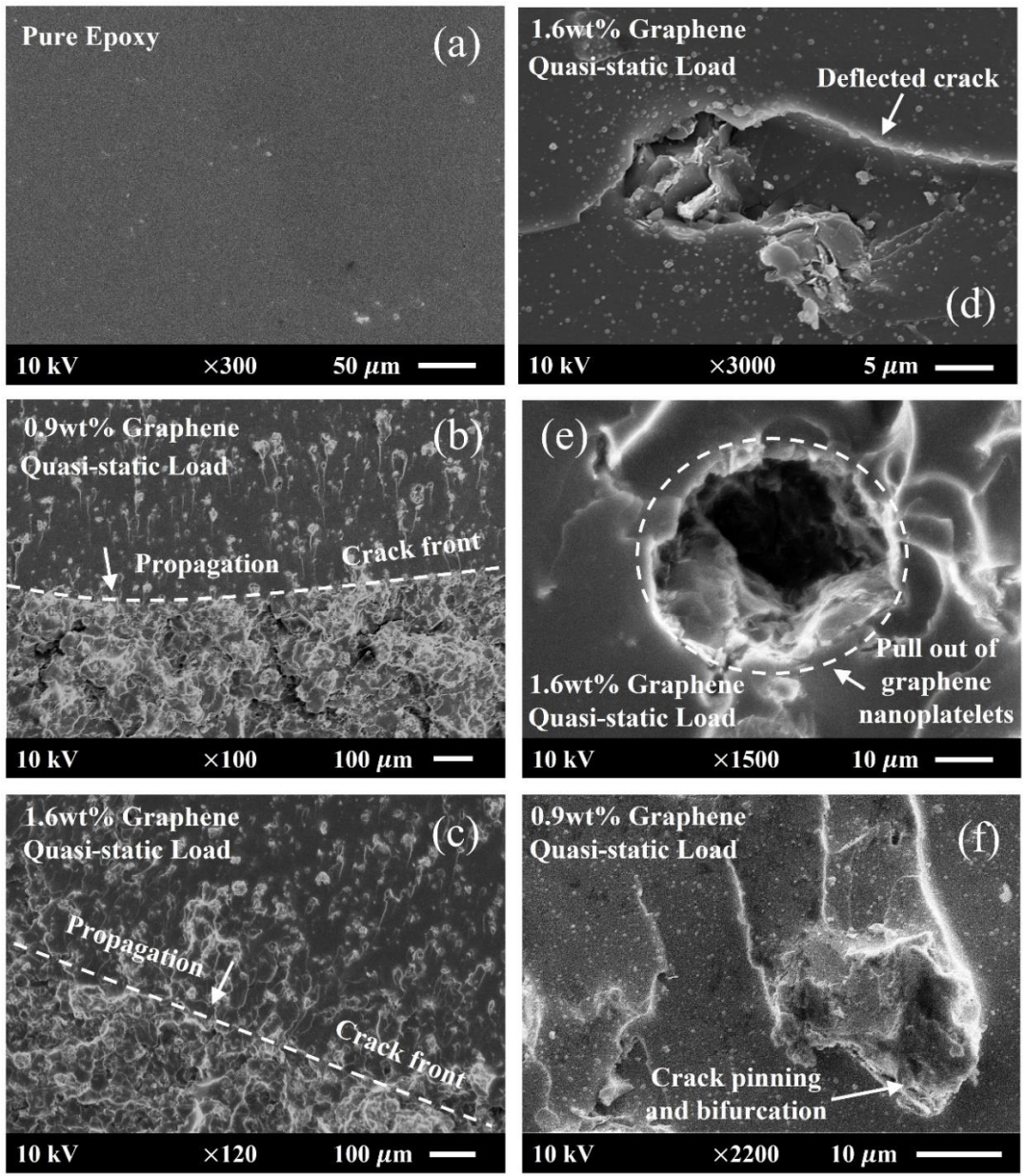

**Figure 5.** Fracture surfaces of the SENB specimens with different graphene contents after quasi-static failure: (a) pure epoxy with $D = 20$mm; (b) 0.9wt% graphene with $D = 20$mm; (c) 1.6wt% graphene with $D = 20$mm. High magnification images: (d) crack deflection (1.6wt% graphene with $D = 20$mm); (e) graphene nanoplatelets pull-out (1.6wt% graphene with $D = 20$mm); (f) crack pinning and bifurcation (0.9wt% graphene with $D = 20$mm).

### 3.3.2. *Fatigue (cyclic) condition*

Further investigation was conducted on the surface morphologies of the SENB specimens after fatigue failure to understand the mechanisms of fatigue crack growth in graphene polymer nanocomposites and related size effects. As shown in Figure 6a-b for the fracture surfaces of pure epoxy, river marks parallel to the cyclic crack propagation direction were observed for all the investigated specimen sizes. This experimental phenomenon completely contradicts the fatigue fracture morphology in metals typically characterized by beach marks perpendicular to the cyclic crack propagation direction [84]. This significant difference can be explained by the fact that the cyclic crack front propagates on different planes in polymeric materials. As clearly illustrated in Figure 6f, a higher magnification of the image shows the river marks on the fracture surface of pure epoxy. On the other hand, the fatigue fracture surfaces of graphene nanocomposites with different specimen sizes do not exhibit river marks as exemplified in Figure 6c-d for 1.6wt% graphene nanocomposites. The existence of the graphene nanoplatelets in a thermoset polymer can definitely deflect the cyclic crack propagation, leading to a tortuous fracture path (Figure 6e) and improving the fatigue lifetime as a consequence, as similarly shown in Section 3.3.1 for quasi-static loading condition and also reported by many studies in the literature (*e.g.*, [85], etc.).

Regarding the size effect on the morphological characteristics of fatigue fracture surfaces of graphene nanocomposites, the specimen with a larger size typically features more pronounced marks compared to the specimen with a smaller size. This can be clearly seen in Figure 6a-d for both investigated pure epoxy and 1.6wt% graphene nanocomposite. The reason can be associated with the rate of the cyclic crack propagation since a smaller specimen has a longer fatigue lifetime to failure as discussed in Section 3.2.

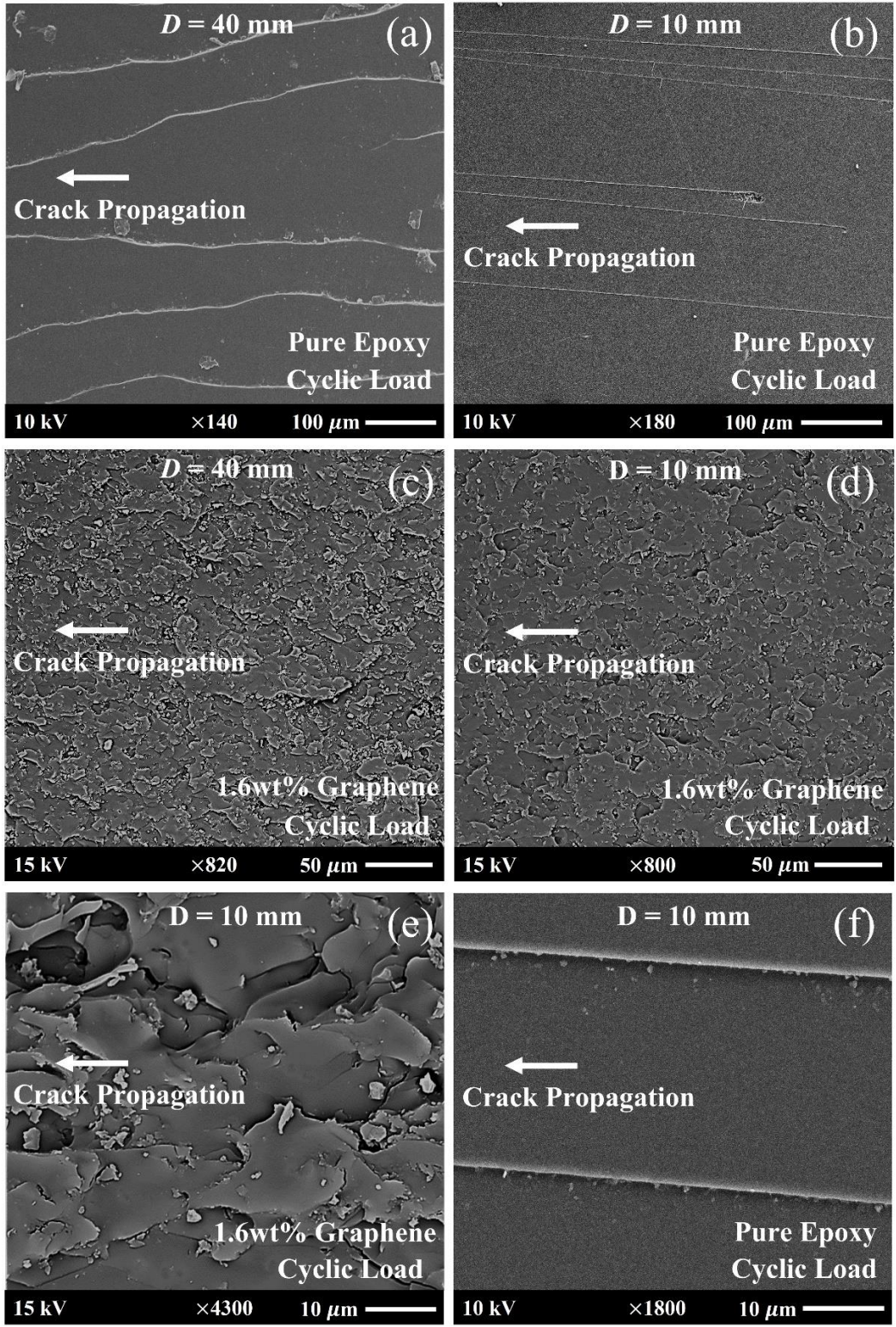

**Figure 6.** Fracture surfaces of the SENB specimens with different sizes and graphene contents after fatigue failure: (a) pure epoxy with $D = 40$mm; (b) pure epoxy with $D = 10$mm; (c) 1.6wt% graphene nanocomposite with $D = 40$mm; (d) 1.6wt% graphene nanocomposite with $D = 10$mm; (e) high magnification image of 1.6wt% graphene nanocomposite with $D = 10$mm; (f) high magnification image of pure epoxy with $D = 10$mm.

## 4. Analyses of Fracture Behavior by Size Effect Method

The previous section shows some fracture morphologies, *e.g.*, discontinuous cracking due to micro-crack pinning, tortuous cracking due to micro-crack deflection, etc. induced by the addition of graphene nanoplatelets. This leads to the formation of a non-linear Fracture Process Zone (FPZ) occurring in the presence of a stress-free crack. The size of this FPZ is generally not negligible compared to the specimen size as also often reported for various engineered materials [86-99]. Consequently, such a fracture behavior and its scaling associated with a given structural geometry, cannot be described by the classical Linear Elastic Fracture Mechanics (LEFM), which neglects the FPZ and treats as a mathematical point in front of the crack tip. To properly capture the scaling of the fracture behavior of graphene nanocomposites, a characteristic and finite length scale related to the materials properties (*e.g.*, fracture energy, material strength, fatigue threshold, etc.) must be introduced to account for the effects of a finite and non-negligible FPZ ahead of the crack tip [62,63]. Details are given in the following sections.

*4.1. Quasi-static loading condition*

*4.1.1. Size effect law (SEL) for quasi-static fracture*

The fracture tests on graphene nanocomposites can be analyzed by leveraging an equivalent LEFM approach to consider the existence of a finite FPZ size ahead of the crack tip, induced by different damage characteristics (*e.g.*, pulling out of graphene nanoplatelets, crack deflection due to graphene nanoplatelets, etc.) as shown in Section 3.3. This method has been shown in the literature to successfully characterize fracture properties of various quasi-brittle materials (*e.g.*, concrete, ceramics, rocks, sea ice, fiber-reinforced polymers, wood, bone, nacre, etc.) [62,63]. In the framework of equivalent LEFM, an equivalent crack length $a_{eq} = a_0 + c_{f,q}$, including an initial crack length ($a_0$) and an effective FPZ length ($c_{f,q}$) in quasi-static loading condition (Figure 7), is considered in the formulation. Accordingly, the fracture energy of the material can be written in the following expression:

$$G_f(\alpha_{eq}) = G_f(\alpha_0 + c_{f,q}/D) = \frac{\sigma_{Nc}^2 D}{E^*} g(\alpha_0 + c_{f,q}/D) \qquad (1)$$

where $\alpha_0 = a_0/D$ is the normalized initial crack length, $\alpha_{eq} = \alpha_0 + c_{f,q}/D$ is the normalized effective crack length, $\sigma_{Nc} = 3PL/(2tD^2)$ is the nominal strength, $E^* = E$ for plane stress condition and $E^* = E/(1-v^2)$ for plane strain condition, $v$ is the Poisson's ratio, $g(\alpha_{eq})$ is the dimensionless energy release rate, and $G_f$ is the fracture energy. It is worth mentioning here that Eq. ( 1 ) describes the peak load condition only if the specimen has a positive geometry

(i.e., $g' > 0$ where $g'$ is the first derivative of $g$) [62,63]. The expression of $g(\alpha_0 + c_{f,q}/D)$ can be expanded by leveraging the Taylor series expansion at $\alpha_0$ and only keeping the linear term of the expansion, Eq. ( 1 ) changes to the following form:

$$G_f(\alpha_{eq}) = \frac{\sigma_{Nc}^2 D}{E^*}\left[g(\alpha_0) + \frac{c_{f,q}}{D}g'(\alpha_0)\right] \qquad (2)$$

where $c_{f,q}$ is the key, not considered in the classical LEFM, to physically capture the formation of the FPZ in front of the crack tip and properly estimate the size-independent fracture energy of the material.

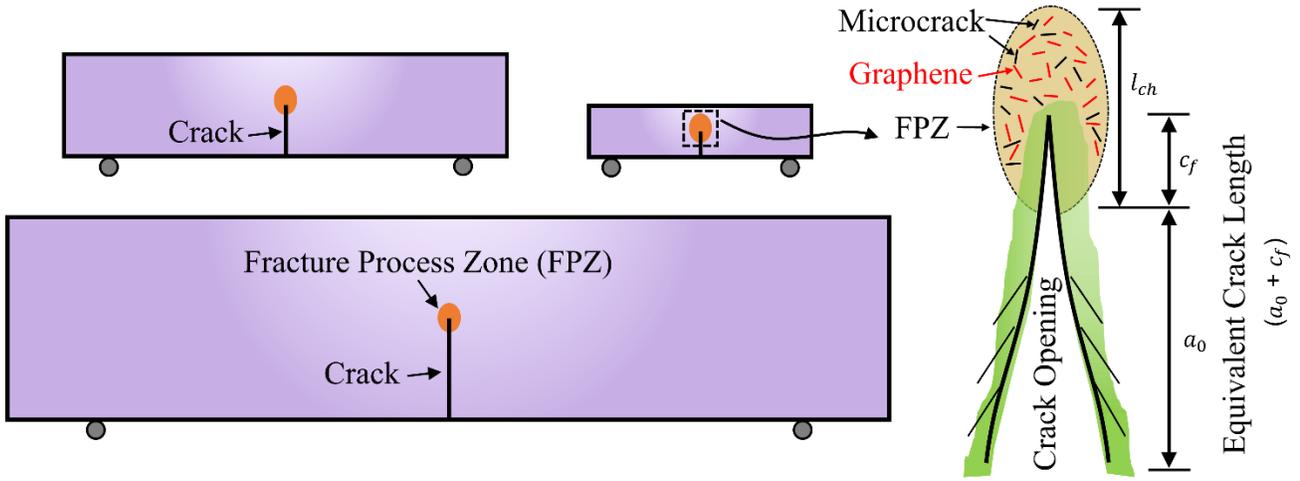

**Figure 7.** Schematic representation of the Fracture Process Zone (FPZ) in front of the crack tip in graphene nanocomposites with different specimen sizes. The equivalent crack length $a_{eq} = a_0 + c_f$, leading to the crack tip stress field following the classical LEFM, where $c_f$ is proportional to the Erwin's characteristic length $l_{ch} = G_f E^*/f_t^2$ and $f_t$ is the material strength.

*4.1.2. Estimation of quasi-static fracture properties*

To obtain the values of $G_f$ and $c_{f,q}$ for the investigated graphene nanocomposites, Eq. ( 2 ) can be rearranged into the following form to fit the experimental data of different specimen sizes for each graphene content:

$$\frac{1}{g'(\alpha_0)\sigma_{Nc}^2} = \frac{g(\alpha_0)}{g'(\alpha_0)G_f E^*}D + \frac{c_{f,q}}{G_f E^*} \qquad (3)$$

$$Y = \frac{1}{G_f E^*}X + \frac{c_{f,q}}{G_f E^*} = AX + B \qquad (4)$$

where $Y = (g'(\alpha_0)\sigma_{Nc}^2)^{-1}$, $X = g(\alpha_0)D/g'(\alpha_0)$, $A = (G_f E^*)^{-1}$, and $B = c_{f,q}/(G_f E^*)$. The functions of $g(\alpha_0)$ and its derivative $g'(\alpha_0)$ for the investigated SENB specimen can be found in a recent publication by Qiao and

Salviato [86] by rearranging the equations in ASTM D5045-99 [73], or can be obtained through the Finite Element Analyses (FEA) as often conducted in the literature (*e.g.*, [97], etc.). By performing a linear regression with Eq. ( 3 ) or ( 4 ) for all the graphene contents as shown in Figure 8, the parameters $A$ and $B$ corresponding to the fracture properties can be obtained for each graphene content. It is interesting to note that the slope $A$ of the regression curve decreases as the graphene content increases. This is an indication of increased fracture energy by the addition of graphene nanoplatelets, because the parameter $A$ is inversely proportional to material fracture energy and elastic modulus ($G_f E^*$) where the elastic modulus shows almost no change for the investigated graphene contents as reported in a recent publication by the authors [25]. On the other hand, as shown in Figure 8, the intercept $B$ does not pass the origin of the coordinate for all the cases, but exhibits a higher value for a higher graphene content. This phenomenon shows an increasing size of the FPZ with the addition of graphene nanoplatelets since a regression line must pass through the origin for a material with a negligible FPZ size as assumed by the LEFM. This fact is also consistent with the load-displacement curves of graphene nanocomposites, showing increased pseudo-ductility before final failure by increasing the graphene content in comparison with that of pure epoxy (Figure 2).

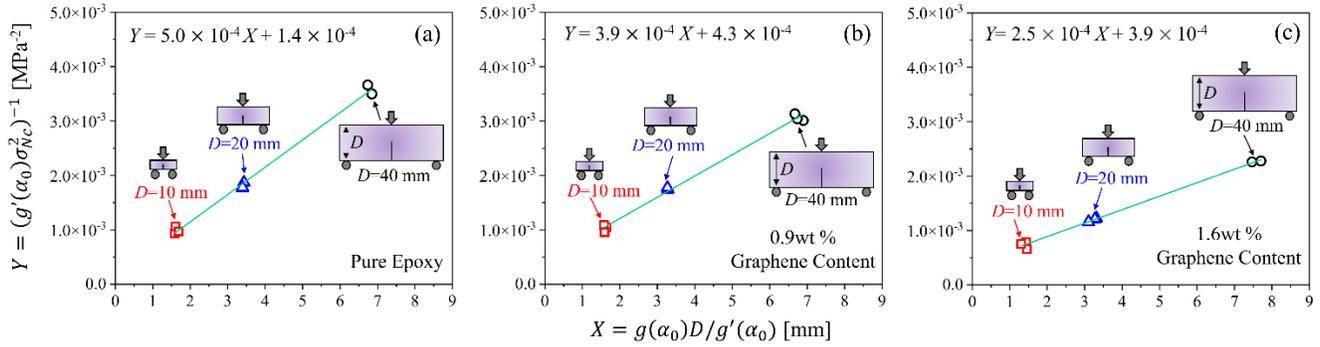

**Figure 8.** Fitting of experimental data of quasi-static fracture tests via Eq. ( 4 ) for the specimens with different sizes and graphene contents.

Having calculated the parameters $A$ and $B$, the effective FPZ length ($c_{f,q}$) for each graphene content can be obtained ($c_{f,q} = B/A$) as plotted in Figure 9a. In this figure, the addition of graphene nanoplatelets leads to a larger $c_{f,q}$ showing about 1.10 mm for 0.9wt% graphene content and 1.59 mm for 1.6wt% graphene content (about 6 times larger than the effective FPZ length of pure epoxy). This aspect is of utmost importance for an accurate calculation of the fracture energy, while the LEFM assuming a negligible FPZ seems reasonable for the pure epoxy, this is not true for graphene-nanomodified specimens which show a FPZ size about one order of magnitude larger and not negligible compared to the specimen width. Thanks to the estimated $c_{f,q}$ value for each graphene content, the Mode

I fracture energies of graphene polymer nanocomposites then can be properly calculated for each graphene content. As plotted in Figure 9b, the Mode I fracture energy increased from 0.88 N/mm for pure epoxy, 1.06 N/mm for 0.9wt% graphene content, to 1.69 N/mm for 1.6wt% graphene content. About 93% improvement on the fracture energy was achieved by adding 1.6wt% graphene nanoplatelets in an epoxy. Thanks to the characteristic length scale associated with the FPZ size, SEL (Eq. ( 4 )) provides the same estimate of the fracture energy for all the specimen sizes (Figure 9b), otherwise the use of the LEFM can lead to the fracture energy strongly depending on the specimen size as previously reported by the authors [25].

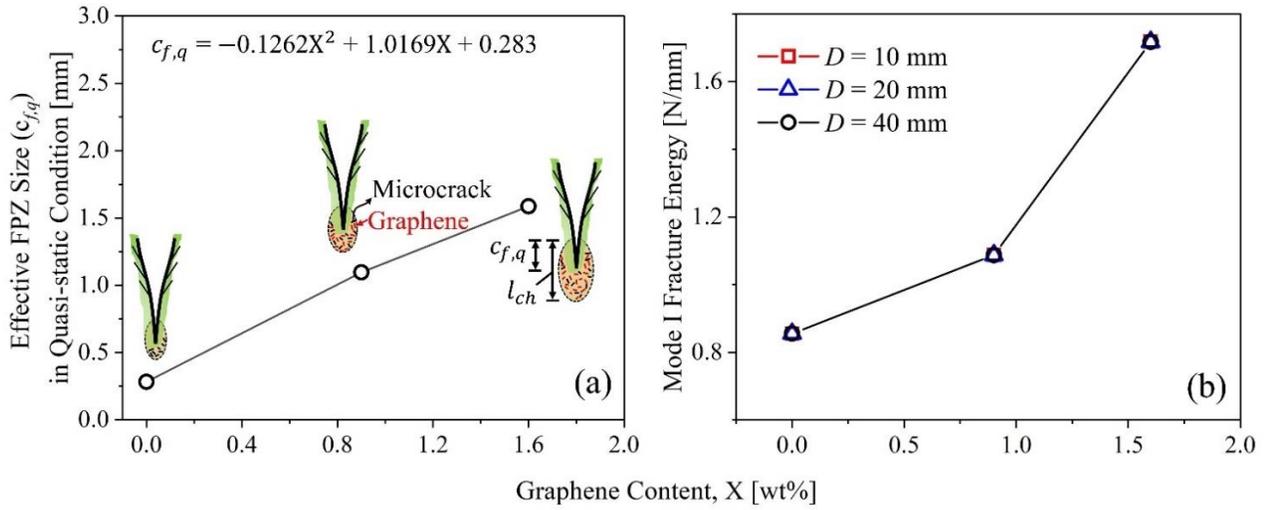

**Figure 9.** (a) Effective FPZ length as a function of graphene content in quasi-static condition; (b) the Mode I fracture energies of different graphene contents estimated from the SEL (Eq. ( 4 )) for different sizes.

### 4.1.3. Scaling of quasi-static fracture behavior

Having known the fracture properties of graphene nanocomposites, the scaling of the fracture behavior can be obtained for a given normalized crack length. By rearranging Eq. ( 2 ), the nominal or structural strength can be written in the following expression:

$$\sigma_{Nc} = \sqrt{G_f(\alpha_{eq})E^*/[Dg(\alpha_0) + c_{f,q}g'(\alpha_0)]} = \sigma_0/\sqrt{1 + D/D_0} \qquad (5)$$

where $\sigma_0 = G_f(\alpha_{eq})E^*/[c_{f,q}g'(\alpha_0)]$ and $D_0 = c_{f,q}g'(\alpha_0)/g(\alpha_0)$. Considering the case that the normalized initial crack lengths for different specimens and sizes from the experiments were not the same, one can use the following equation to convert the experimental data to a desired case (*e.g.*, $a_0 = 0.5D$, etc.) by imposing the same fracture energies for both cases:

$$\frac{\sigma_{Nc,exp}^2 D}{E^*}\left[g(\alpha_{0,exp}) + \frac{c_{f,q}}{D}g'(\alpha_{0,exp})\right] = \frac{\sigma_{Nc,desire}^2 D}{E^*}\left[g(\alpha_{0,desire}) + \frac{c_{f,q}}{D}g'(\alpha_{0,desire})\right] \quad (6)$$

Eq. (6) can be further reduced into the following expression for the correlation between $\sigma_{Nc,exp}$ and $\sigma_{Nc,desire}$:

$$\sigma_{Nc,desire} = \sigma_{Nc,exp}\sqrt{[Dg(\alpha_{0,exp}) + c_{f,q}g'(\alpha_{0,exp})]/[Dg(\alpha_{0,desire}) + c_{f,q}g'(\alpha_{0,desire})]} \quad (7)$$

Accordingly, the adjusted experimental data through Eq.(7) for a given normalized crack length ($\alpha_{0,desire} = 0.5D$) and the fitting by the SEL were shown in Figure 10. In addition, the predictions by the SEL for several other normalized crack lengths were also plotted in Figure 10. In this figure, the critical nominal or structural strength was plotted as a function the specimen width $D$ in a double logarithmic scale. In such a graph, the scaling behavior predicted by the LEFM is represented by a line of slope -1/2, whereas a horizonal line represents the case of no scaling following the stress-based failure criteria. In other words, the foregoing two extremes (*i.e.*, the LEFM and plastic asymptotes) represent a typical brittle behavior and a ductile behavior respectively. The intersection between two asymptotes, corresponding to $D = D_0$ as often called the transitional size [62,63], represents a transitional region between brittleness and ductility (*i.e.*, quasi-brittleness).

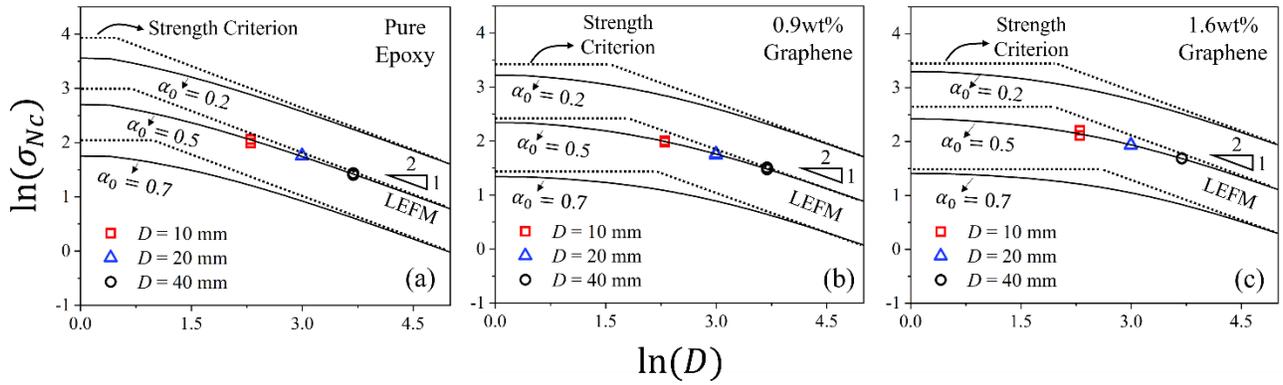

**Figure 10.** Experimental data and size scaling curves of quasi-static structural strength for the specimens with different normalized initial crack lengths ($\alpha_0$) estimated by using Eq. (5): (a) pure epoxy; (b) 0.9wt% graphene content; (c) 1.6wt% graphene content.

In Figure 10a, for a given normalized initial crack length ($\alpha_0 = 0.5D$), the structural strengths of the pure epoxy system with the investigated sizes are closely located at the LEFM asymptote (dashed lines) in spite of a little deviation from the SEL for a small specimen size $D = 10$ mm. This aspect shows a good description of the quasi-static fracture scaling in pure epoxy by the LEFM for the range of the investigated sizes, and confirms that the FPZ size has a negligible effect on pure epoxy with a sufficiently large specimen size. In this context, the LEFM approach can be used as suggested by ASTM D5045-99 [73], not considering the FPZ ahead of the crack tip.

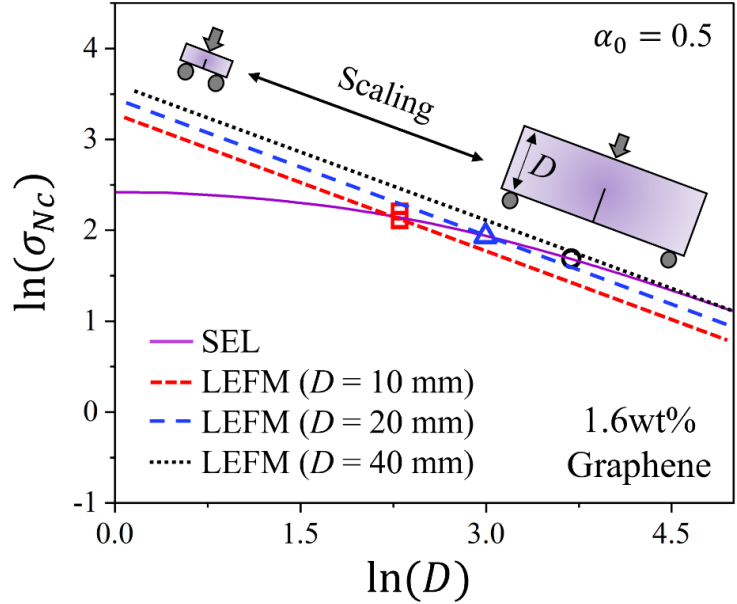

**Figure 11.** Scaling of the fracture behavior of 1.6wt% graphene nanocomposites predicted by the SEL and the LEFM methods. Note, the SEL was based on the experimental data of different specimen sizes, whereas the LEFM was based on the individual specimen size.

However, this is not the case for graphene nanocomposites with $\alpha_0 = 0.5D$ as shown in Figure 10b-c. The experimental data has a significant deviation from the LEFM (dashed lines) for both 0.9wt% and 1.6wt% graphene contents, being more pronounced for a higher graphene content and a smaller specimen size. This phenomenon is attributed to the increased FPZ size in graphene nanocomposites compared to the structure size, which makes the non-linear effects caused by the damage in front of the crack tip not negligible. For sufficiently small specimens, the FPZ affects the failure behavior of graphene nanocomposites and causes a significant deviation from the fracture scaling predicted by the LEFM. When increasing specimen sizes of graphene nanocomposites, the effects of the FPZ become less significant, thus being closely captured by the LEFM. Thanks to the SEL, the experimental data of graphene nanocomposites, locating at the transitional region with two extremes (*i.e.*, stress-driven failure by a horizontal asymptote and energy-driven fracture by the LEFM asymptote), can be well captured as shown in Figure 10. Moreover, it is interesting to notice from Figure 10 that both the addition of graphene nanoplatelets and the normalized initial crack length can shift the failure behavior towards the ductile region (*i.e.*, increasing $D_0$) due to the increased ratio of the FPZ size to ligament length (*i.e.*, $D - a_0$).

The foregoing results are important for graphene nanocomposites and nanoparticle-reinforced polymers in general. As the experimental data and the analytical analyses show, the LEFM does not always provide an accurate method to extrapolate the structural strength of larger structures from lab tests on small-scale specimens, in particular for the specimen sizes belonging to the transitional region. As illustrated in Figure 11 by taking an example of 1.6wt% graphene content and a given normalized initial crack length ($\alpha_0 = 0.5$), the use of the LEFM in such cases can lead to a significant underestimation of structural strength. This hinders the structural performance optimization for several engineering fields (*e.g.*, automobiles, aviation, etc.). On the other hand, when using the LEFM to predict the structural strength for smaller length scales, the predicted results can be largely overestimated (Figure 11). However, the SEL or other models considering a characteristic length scale can be used to provide a better description of fracture scaling from brittle, transitional, and ductile regions in nanocomposites and other quasi-brittle materials.

*4.2. Fatigue (cyclic) loading condition*

4.2.1. *Paris-Erdogan curves*

Before discussing the SEL used for fatigue fracture tests in the next section, it is important to firstly show the Paris-Erdogan curves [100] of graphene nanocomposites with different specimen sizes, constructed by using the cyclic evolution of the normalized equivalent crack length ($\alpha_{eq} = a_{eq}/D$) (Figure 3). As shown in Figure 12, the Stress Intensity Factor (SIF) amplitude for fatigue ($\Delta K_{eq}$) was plotted with the normalized equivalent crack length per cycle ($d\alpha_{eq}/dN$) in a double logarithmic scale. It is worth mentioning here that the foregoing equivalent crack length includes the cyclic FPZ (See Appendix. 1).

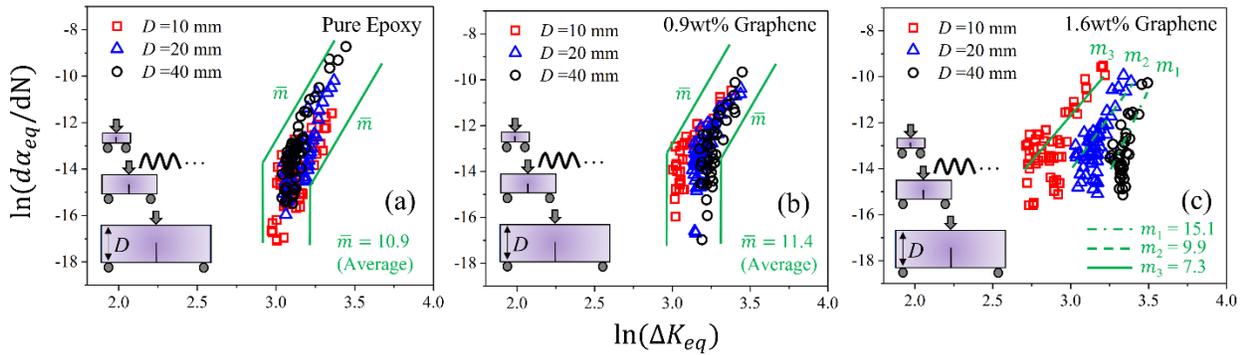

**Figure 12.** Experimental Paris-Erdogan curves for geometrically-scaled graphene nanocomposites. Note that the cyclic growth of the equivalent crack length was estimated by using the compliance method (Appendix 1).

In Figure 12, the fatigue thresholds for different specimen sizes start to exhibit a noticeable difference as graphene content increases, in particular for 1.6wt% graphene content as can be clearly seen from Figure 12c. While the calculated equivalent crack length in fatigue case based on the compliance method (Appendix .1) considers the cyclic FPZ size, the foregoing phenomenon is an indication that the cyclic FPZ size at the fatigue initiation has a different development for different specimen sizes, being much more pronounced for a higher graphene content. To know the full development of the cyclic FPZ at fatigue initiation, the SEL can be similarly used, which is the topic of Section 4.2.2. Regarding the slope of the Paris-Erdogan curve, representing the stable cyclic crack propagation region, for different sizes and graphene contents, this aspect will be commented in Section 4.2.6.

*4.2.2.  Size effect law (SEL) for fatigue fracture initiation*

Similarly, the equivalent LEFM approach can also be utilized to analyze the fatigue fracture tests on graphene nanocomposites for the measurement of their fatigue fracture properties. The cyclic fracture condition can be written in the following form:

$$\Delta G_{th}(\alpha_{eq}) = \Delta G_{th}(\alpha_0 + c_{f,c}/D) \approx \frac{(\sigma_{n,max}^2 - \sigma_{n,min}^2)D}{E^*}\left[g(\alpha_0) + \frac{c_{f,c}}{D}g'(\alpha_0)\right] \quad (8)$$

where $\Delta G_{th} = \Delta G_{max} - \Delta G_{min}$ is the energy amplitude of fatigue initiation, $c_{f,c}$ is the effective cyclic FPZ length, $\sigma_{n,max}$ and $\sigma_{n,min}$ are the nominal stresses $\sigma_n = 3PL/(2tD^2)$ calculated by means of the forces (*i.e.*, $P_{max}$ and $P_{min}$) at the peak and bottom of an applied cyclic load. By using the stress ratio $R = \sigma_{n,min}/\sigma_{n,max}$, Eq. ( 8 ) can be simplified into the following expression:

$$\Delta G_{th}(\alpha_{eq}) = \Delta G_{th}(\alpha_0 + c_{f,c}/D) \approx \frac{R^*(\Delta\sigma_{n,th})^2 D}{E^*}\left[g(\alpha_0) + \frac{c_{f,c}}{D}g'(\alpha_0)\right] \quad (9)$$

where $R^* = (1 + R)/(1 - R)$, $\Delta\sigma_{n,th} = \sigma_{n,max} - \sigma_{n,min}$ is nominal stress amplitude of fatigue initiation, and other symbols have the same meaning as defined in quasi-static case.

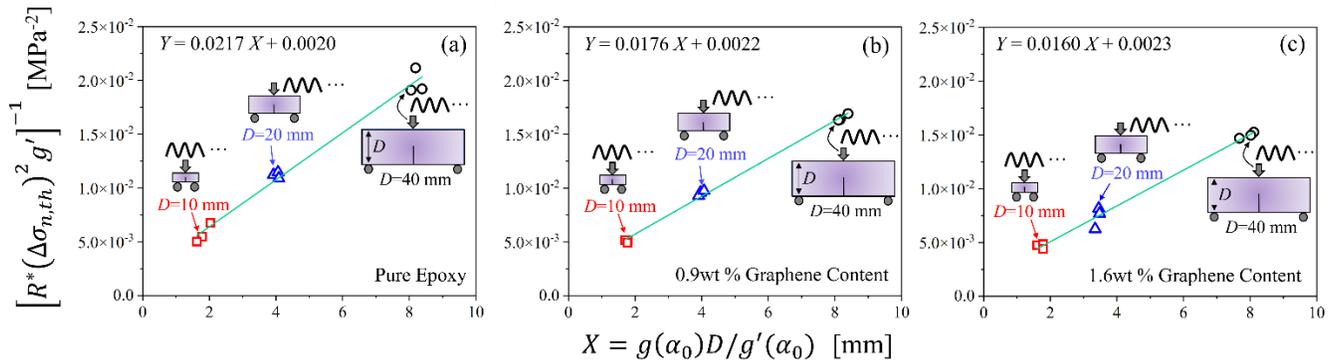

**Figure 13.** Fitting of experimental data of fatigue fracture tests via Eq. ( 11 ) for the specimens with different sizes and graphene contents.

### 4.2.3. Estimation of properties of fatigue fracture

Now, the fatigue fracture properties, $\Delta G_{th}$ and $c_{f,c}$, can be determined by the regression analysis of the experimental data on geometrically scaled specimens as similarly conducted in quasi-static case (Section 2.2.1). For clarity, the expression used for a linear regression analysis of fatigue case was written here:

$$\frac{1}{R^*(\Delta\sigma_{n,th})^2 g'(\alpha_0)} = \frac{g(\alpha_0)}{g'(\alpha_0)\Delta G_{th} E^*} D + \frac{c_{f,c}}{\Delta G_{th} E^*} \quad (10)$$

$$Y = \frac{1}{\Delta G_{th} E^*} X + \frac{c_{f,c}}{\Delta G_{th} E^*} = A_f X + B_f \quad (11)$$

where $Y = \left[1/\left(R^*(\Delta\sigma_{n,th})^2 g'(\alpha_0)\right)\right]^{-1}$ and $X = g(\alpha_0)D/g'(\alpha_0)$.

As plotted in Figure 13, the experiment data of the fatigue size effect tests on different graphene contents was fitted by leveraging the linear regression analyses of Eq. ( 11 ) (*i.e.*, SEL). In this figure, the regression curves do not pass through the origin, confirming that the formation of the non-negligible FPZ also happens at fatigue initiation before fatigue crack propagation for all the investigated graphene contents. In addition, the slope of the regression curve gradually decreases as the graphene content increases, whereas the intersection of the curve exhibits an opposite trend. This was similarly observed from the fitting of the experiment data of quasi-static tests (Figure 8). Moreover, a lower slope $A_f$ and a higher intersection $B_f$ due to the addition of graphene nanoplatelets, as shown in Figure 13, mean a higher energy amplitude required for fatigue crack initiation ($\Delta G_{th}$) and a larger effective cyclic FPZ size ($c_{f,c}$) ahead of the crack tip.

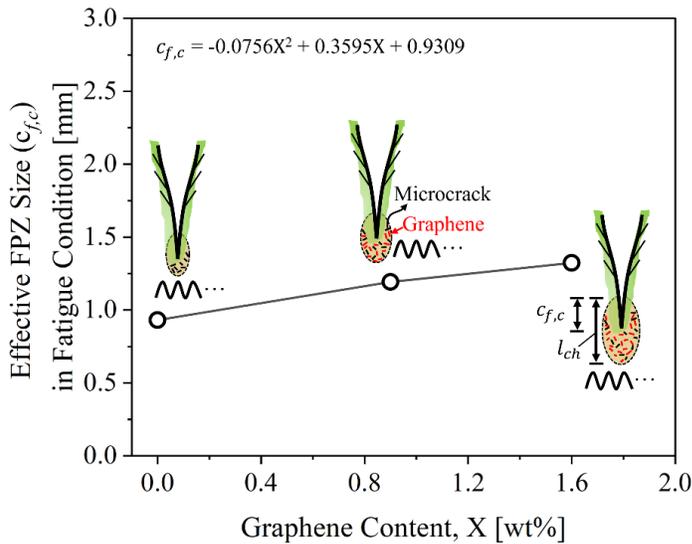

**Figure 14.** Effective FPZ length as a function of graphene content in fatigue condition.

For each graphene content, the effective cyclic FPZ length can be obtained through $c_{f,c} = B_f/A_f$. As shown in Figure 13, $c_{f,c}$ increases from 0.9 mm for pure epoxy to 1.4 mm for 1.6wt% graphene content. This explained the improved fatigue lifetime to failure with the graphene nanomodification (Figure 4) due to increased fracture ductility ahead of the crack tip. It is worth emphasizing here that the $c_{f,c}$ estimated from the fatigue SEL method (Eqs. ( 10 ) and ( 11 )) for each graphene content should be associated with fully developed FPZ size in fatigue condition. The entire evolution of the cyclic FPZ up to the final fatigue failure can only be obtained by means of a numerical fatigue modeling with a physics-informed constitutive model, which is considered for future publications

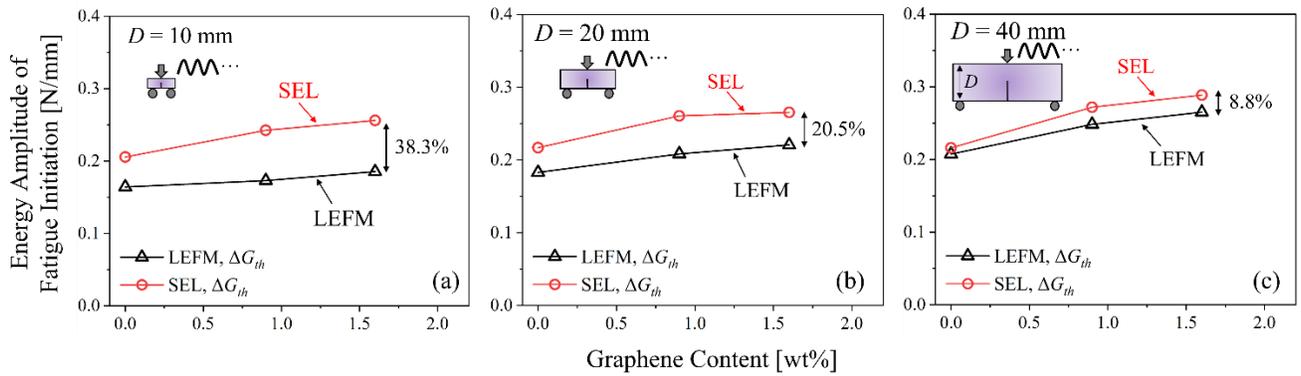

**Figure 15.** Energy amplitudes of fatigue initiation for different graphene contents estimated from the SEL (Eq. ( 9 )) for different specimen sizes. This figure also compares the calculation of the energy amplitudes of fatigue initiation between the LEFM and the SEL.

On the side of energy amplitude of fatigue initiation, a comparison on the results obtained from the SEL and the LEFM was plotted in Figure 15. Similar to the quasi-static case [25], for a give specimen size, the difference between the SEL and the LEFM increases as the graphene content increases due to an enlarged cyclic FPZ size induced by the addition of graphene nanoplatelets. Accordingly, the LEFM estimation of $\Delta G_{th}$ shows size dependency, but this is not the case by using the SEL. For instance, the underestimation of the LEFM for 1.6wt% graphene content can be about 40%, 21%, and 10% for the specimen with $D$ = 10 mm, 20 mm, and 40 mm respectively. The underestimation decreases as the structure size increases also due the negligible ratio of the cyclic FPZ to specimen dimension. In addition, same phenomenon as observed from the quasi-static case [25], while graphene nanocomposites do not show improvement if the LEFM is used for a smaller specimen size, this is not the case for a larger specimen size showing improved $\Delta G_{th}$ after graphene nanomodifications. This may also explain the inconsistent data of fatigue fracture improvement by the addition of different nanofillers in the literature.

*4.2.4. Scaling of fatigue fracture initiation*

By rearranging Eq. ( 9 ), the nominal stress amplitude of fatigue initiation can be written in the following expression:

$$\Delta\sigma_{n,th} = \sqrt{\Delta G_{th}(\alpha_{eq})E^*/[(Dg(\alpha_0) + c_{f,c}g'(\alpha_0))R^*]} = \Delta\sigma_{0,c}/\sqrt{1 + D/D_{0,c}} \qquad (12)$$

where $\Delta\sigma_{n,th} = \Delta G_{th}(\alpha_{eq})E^*/[c_{f,c}g'(\alpha_0)R^*]$ and $D_{0,c} = c_{f,c}g'(\alpha_0)/g(\alpha_0)$. It is worth mentioning here that a characteristic length scale $c_{f,c}$ plays a pivotal role in the transition from stress-dominated fatigue (*i.e.*, stress-life behavior) to energy-dominated fatigue (*i.e.*, crack growth rate versus SIF) as the structure size or initial crack length increases. Similar to the scaling of quasi-static fracture in Section 4.1.3, the nominal stress of fatigue initiation $\Delta\sigma_{n,th}$ for a desired normalized initial crack length (*e.g.*, $a_0 = 0.35D$, etc.) can be obtained by using the following equation assuming the same $\Delta G_{th}$ for the specimens with different initial crack lengths:

$$\Delta\sigma_{N,th,desire} = \Delta\sigma_{N,th,exp}\sqrt{[Dg(\alpha_{0,exp}) + c_{f,c}g'(\alpha_{0,exp})]/[Dg(\alpha_{0,desire}) + c_{f,c}g'(\alpha_{0,desire})]} \qquad (13)$$

By adjusting the experimental data through Eq. ( 13 ) for a normalized initial crack length ($a_0 = 0.35D$) as an example, the results and the fitting by the SEL (Eq. ( 12 )) were plotted together in Figure 16. In this figure, the quantity related to the nominal stress amplitude of fatigue initiation $\Delta\sigma_{n,th}\sqrt{R^*}$ was plotted as a function of the specimen width $D$ in a double logarithmic scale. As shown in Figure 16a, the applied nominal cyclic stress amplitude to trigger a stable fatigue crack propagation can be successfully described by the LEFM for a graphene nanocomposite with a sufficiently large specimen. However, it does not happen for a smaller specimen size since the value of $\Delta\sigma_{n,th}\sqrt{R^*}$ shows a deviation from the LEFM as can be seen from Figure 16a. Thanks to the consideration of the cyclic FPZ in Eq. ( 12 ), the scaling of fatigue initiation of graphene nanocomposites can be well captured, which shows the transition from energy-dominated fatigue to stress-dominated fatigue as the size increases. For different graphene contents as shown in Figure 16a , all the cases including pure epoxy are located in the transition region exhibiting minor differences. Notwithstanding this, the addition of graphene nanoplatelets can slightly increase the $\Delta\sigma_{n,th}\sqrt{R^*}$ as the specimen size increases. However, for a smaller size, the prediction of the fatigue SEL shows a lower asymptote for a higher graphene content, indicating that the addition of graphene nanoplatelets can reduce the fatigue initiation for a sufficiently small specimen size. A better understanding of graphene nanomodification on the scaling of fatigue fracture requires a comprehensive study of computational modeling which is considered for future publications.

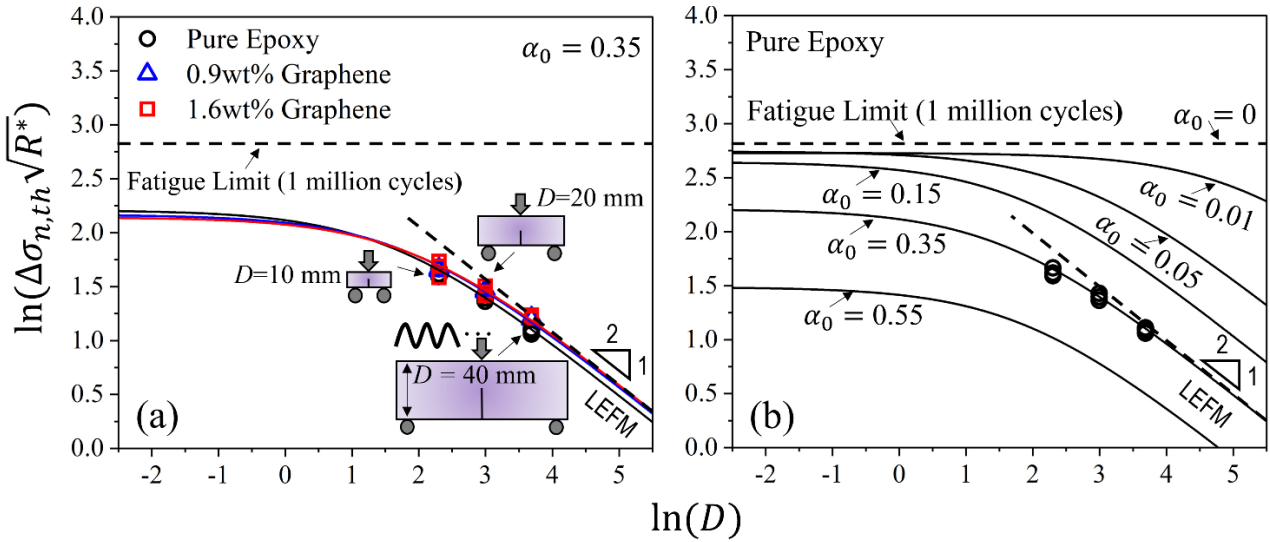

**Figure 16.** Experimental data and size scaling curves of fatigue initiation for different graphene contents estimated by using Eq. ( 12 ); (b) Size scaling curves for pure epoxy with different normalized initial crack lengths ($\alpha_0$) as an example. This figure shows the transition between fatigue crack propagation to fatigue initiation by reducing the specimen size or initial crack length.

On the other hand, it is interesting to check the effects of initial crack length on the fatigue initiation as similarly conducted for quasi-static loading condition in Section 4.1.3. By taking an example of pure epoxy considering the similar size effect curves for graphene nanocomposites in Figure 16a, $\Delta\sigma_{n,th}\sqrt{R^*}$ as a function of the specimen size for different normalized initial crack lengths were plotted Figure 16b by means of Eq. ( 12 ). As can be noted from the figure, the size scaling curve of fatigue initiation becomes milder as the normalized initial crack length decreases, and eventually shows a horizontal line for an unnotched specimen ($a_0 = 0$). This horizontal line was obtained from the fatigue testing on unnotched specimens of pure epoxy up to 1 million cycles without failure [87]. The foregoing aspect indicates that, by decreasing the ratio of the initial crack length to specimen size, the scaling of fatigue fracture transits from type II (pre-notched structures) energetical scaling to type I (un-notched structures) energetical-statistical scaling, similar to the effects of initial crack length on quasi-static scaling [101]. Moreover, it is worth mentioning here that the fatigue limit (Figure 16) can be even lower since the fatigue with sufficiently high number of cycles were not investigated in this work and also less explored in the literature due to time-consuming fatigue testing. In fact, ultra-high cyclic fatigue behavior of polymer composites was reviewed by Ding *et al.* [102] who showed fatigue failure of polymer composites even after 10 million cycles. This means that the fatigue limit can be lower than the asymptotes of different normalized initial crack lengths, and this aspect is consistent with the observation in the quasi-static transition from type II to type I as reported by Hoover and Bažant [101] for concrete structures – an archetype of quasi-brittle materials.

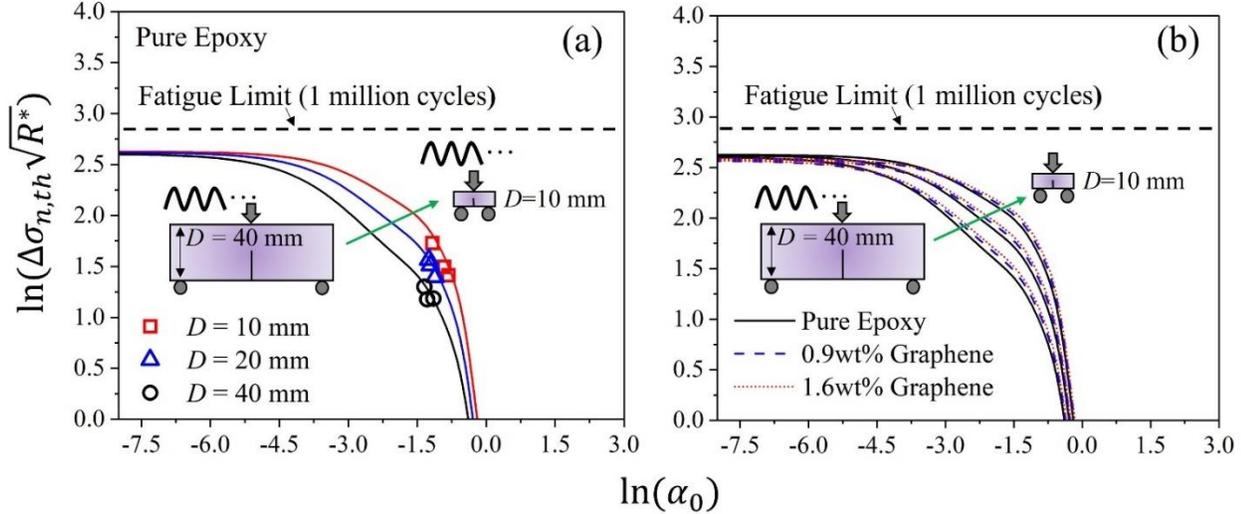

**Figure 17.** (a) Experimental data of pure epoxy and its size-dependent Kitagawa-Takahashi diagram captured by using Eq. ( 12 ). The area inside the envelop represents the safe region for no fatigue crack propagation, whereas the area outside the curve means the fatigue crack growth region; (b) Comparison on the size-dependent Kitagawa-Takahashi diagrams (Eq. ( 12 )) among different graphene contents.

*4.2.5. Stress amplitude for fatigue initiation vs. initial crack length*

In addition, it is worth clarifying an important aspect for fatigue initiation in this study, which was often overlooked in the literature. It is the well-known Kitagawa-Takahashi diagram and its modified version [103,104], which were often used for various quasi-brittle materials to provide a critical envelop for the combination of initial crack length and the cyclic stress amplitude of fatigue initiation. However, such a diagram is size dependent as same as the classical Paris-Erdogan curve [100], which is not a unique fatigue property for different specimen sizes. The reason is due to the non-negligible FPZ size ahead of the crack tip compared to the structure size, as discussed throughout this paper, in fatigue condition.

As exemplified in Figure 17a for the excellent characterization of pure epoxy with different sizes by means of the fatigue SEL (Eq. ( 12 )), the associated nominal stress amplitude, leading to the fatigue crack initiation, was plotted as a function of the normalized initial crack length $a_0$ in a double logarithmic coordinate. In this figure, $\Delta\sigma_{n,th}\sqrt{R^*}$ approaches the fatigue limit (*i.e.*, 1 million cycles of unnotched epoxy without failure as considered in this study) as the initial crack length decreases. This aspect shows the bridging between fatigue crack initiation and propagation. Moreover, the area inside of the critical envelop is considered as the safe region where the initial crack length does not propagate and not lead to fatigue failure, whereas the cases on or outside of the critical envelop represents that the material will experience fatigue failure. In other words, the location far from the critical envelop indicates a short fatigue lifetime, but a longer or infinite fatigue lifetime requires the location close or inside of the critical envelop.

More importantly, the critical envelopes of graphene nanocomposites show remarkable size dependence in the transitional region as shown in Figure 17. The critical envelop expands as the specimen size decreases. The reason is due to the non-negligible cyclic FPZ compared to the specimen size, as shown by the experimental data and the predictions from the fatigue SEL (Eq. ( 12 )). However, two extremes (*i.e.*, sufficiently short and large initial crack lengths) are less affected by the sizes, which is reasonable since the cyclic FPZ sizes have similar effects on both cases. It is also interesting to notice that the critical envelopes slightly expand as graphene content increases as shown in Figure 17b. This aspect confirms the improved fatigue lifetime and $\Delta G_{th}$ with the addition of the investigated graphene nanoplatelets as discussed in previous sections.

### 4.2.6. *Discussion on fatigue fracture propagation rate*

Regarding the fatigue crack propagation in graphene nanocomposites, the representative slopes of the Paris-Erdogan curves, $\bar{m} = 10.9$ and $11.4$, can be used to describe the experimental data of different sizes for pure epoxy and 0.9wt% graphene content respectively as shown in Figure 12a-b. This slope $\bar{m}$ was based on an average value of all the specimen sizes with three experimental data for each specimen size. However, this is not a true for 1.6wt% graphene content, as shown in Figure 12c, the Paris-Erdogan curves exhibit non-negligible difference on the slopes for different specimen sizes. The slope $m = 7.3$ for a smaller specimen size $D = 10$ mm, whereas a larger specimen size $D = 40$ mm has a steeper slope of $m = 15.1$, being about two times larger. The slope $m$ for each specimen size was based on three experimental tests.

The difference on the slopes of Paris-Erdogan curves for a higher graphene content indicates that a crack in a quasi-brittle material may have incomplete self-similar propagation for different specimen sizes. The incomplete self-similarity of crack propagation was actually a mathematical hypothesis by leveraging the Buckingham dimensionless analysis [105] as studied by several researchers (*e.g.*, Barenblatt [106,107], Ritchie [108], Ciavarella [109], etc.) in the past decades. Why was this phenomenon not observed from the fatigue size effect tests of other quasi-brittle materials (e.g., concrete [57-59], sandstone [60], etc.)? The answer requires a comprehensive computational modeling of the cyclic cohesive fracture in materials with different sizes. The related study is going to be published by the authors in an incoming journal article.

5. **Conclusions**

This paper investigated the size scaling effects on the mechanical behaviors and damage characteristics of graphene nanocomposites in both quasi-static and fatigue (cyclic) loading conditions. Based on the results obtained from this study, the following conclusions can be elaborated:

1. in quasi-static loading condition, the addition of 1.6wt% graphene nanoplatelets in pure epoxy can increase the Mode I fracture energy up to 1.69 N/mm, being about 93% higher than that of pure epoxy. Such an enhancement was due to *e.g.*, crack deflection, pinning, bifurcation, and pulling out of graphene nanoplatelets, etc. These damage mechanisms dissipate more energy during crack formation process, thus increasing the quasi-brittleness of graphene-reinforced polymers and improving their fracture resistance;

2. the analysis of quasi-static fracture tests showed that, while the scaling on the structural strength of pure epoxy can be roughly captured by the classical Linear Elastic Fracture Mechanics (LEFM) with $-1/2$ asymptote indicating brittleness, this was not the case for graphene nanocomposites showing a non-linear fracturing scaling with a deviation from the LEFM meaning ductility. This phenomenon was more and more pronounced as graphene content increases;

3. the deviation from the LEFM due to a non-linear scaling is attributed to the Fracture Process Zone (FPZ) size for increasing the graphene content. For pure poxy, the FPZ ahead of the crack tip due to subcritical damage (*e.g.*, crazing [74,110,111], etc.) is very small compared to the investigated specimen sizes. This aspect agrees with the LFEM assuming a negligible FPZ size (a mathematical point) in front of the crack tip. However, for graphene nanocomposites with higher graphene contents and smaller specimen sizes, the FPZ size formed by aforementioned damage mechanisms was large and not negligible compared to the specimen characteristic size, thus causing a non-linear fracture scaling and a deviation from the LEFM;

4. to capture such a complicated scaling in quasi-static case, a characteristic length $c_{f,q}$ related to the FPZ size must be considered into the equation, known as Size Effect Law (SEL) [62,63]. Accordingly, an excellent agreement of the experimental data of graphene nanocomposites, a non-linear fracture scaling from brittle to ductile region, can be achieved by the SEL. The calculated $c_{f,q}$ increased from 0.28 mm for pure epoxy to 1.59 mm for 1.6wt% graphene content in quasi-static case;

5. in fatigue loading condition, the classical Paris-Erdogan curves of pure epoxy constructed by directly measuring the crack length were noticeably affected by the specimen size, showing the differences mainly on the fatigue threshold and the intersection of the curve. The reason was due to the fact the classical Paris-Erdogan curve was based on the LEFM with the assumption of a mathematical point as the damage ahead of the crack tip. This contradicts to the enlarged cyclic FPZ ahead of the crack tip (*e.g.*, micro-cracks at different planes forming river marks, etc.) in pure poxy as similarly discussed for quasi-static case;

6. the foregoing difference on the fatigue threshold (initiation) of pure epoxy with different sizes can be removed by leveraging an analytical model based on an energetic-equivalence framework (Appendix 2), considering the cyclic FPZ ahead of the crack tip, to fit the experimental data of geometrically scaled specimens. However, size-dependent intersections of the Paris-Erdogan curves still exist:

7. this phenomenon can be explained that only the constant cyclic FPZ is taken into the consideration in the foregoing model but not for the cyclic evolution of the FPZ during the fatigue fracture process. By using the compliance method (Appendix 1), the evolution of a characteristic length $c_{f,c}$, related to the cyclic evolution of the FPZ, can be accounted into an equivalent crack length. Consequently, the consistent Paris-Erdogan curves for pure epoxy with different sizes can be achieved;

8. however, the Paris-Erdogan curves of graphene nanocomposites analyzed by the compliance method tend to exhibit differences as the graphene content increases. The most remarkable differences were observed from the case of 1.6wt% graphene content, showing that the fatigue threshold, intersections, and slopes of the Paris-Erdogan curves are not consistent for different specimen sizes;

9. the difference on the fatigue threshold of 1.6wt% graphene nanocomposites with different sizes is mainly attributed to the different cyclic FPZ sizes ahead of the crack tip at fatigue initiation, whereas the different intersections and slopes of the Paris-Erdogan curves are related to size-dependent fatigue crack evolution. The explanation requires a comprehensive study via computational modeling, which is beyond the scope of this study and will be discussed in future publications.

10. by using the proposed SEL endowed with a characteristic length $c_{f,c}$ relating to the cyclic FPZ, the size-independent fatigue properties of graphene nanocomposites can be obtained as similarly discussed for quasi-static case. A weight fraction of 1.6% graphene nanoplatelets can increase the energy amplitude of fatigue initiation up to 0.29 N/mm, being about 32% higher than that of pure epoxy (0.22 N/mm). This value was not dependent on the size, whereas the LEFM without considering the cyclic FPZ led to the size-dependent results as similarly observed from the quasi-static case: 0.185 N/mm, 0.22 N/mm, 0.265 N/mm for the investigated small, medium, and large sizes respectively;

11. the enhanced energy amplitude of fatigue initiation by the addition of graphene nanoplatelets extended the fatigue lifetime to failure. By taking an example of 1.6wt% graphene content, the fatigue lifetime to failure can be improved up to about 34% on average for each specimen size due to the enhanced quasi-brittleness of the material caused by aforementioned damage mechanisms related to graphene nanoplatelets;

12. on the other hand, the SEL analysis of fatigue fracture data showed that the characteristic length $c_{f,c}$ associated with the cyclic FPZ size increased from 0.93 mm for pure epoxy to 1.32 mm for 1.6wt% graphene nanocomposites. This is a confirmation that the fully developed FPZ size in fatigue initiation is larger than that in quasi-static case, as often reported in the literature for various quasi-brittle materials (*e.g.*, [60,112-117], etc.). In addition, the calculated $c_{f,c}$ for pure epoxy is very close to the one from an energetic-equivalence framework (Appendix 2), confirming the validity of the SEL for fatigue analysis;

13. the enlarged FPZ size due to a cyclic loading and a higher graphene content is not negligible compared to the specimen size, which can cause two important aspects. One is the LEFM-based Kitagawa-Takahashi diagram, separating the safe region without failure and the fatigue crack propagation region leading to the final failure. This diagram starts to become size-dependency as the cyclic FPZ increases. The other is the fatigue fracture scaling on the critical energy amplitude of fatigue initiation, also showing a non-linear scaling and a significant deviation from the LEFM as the specimen size decreases or cyclic FPZ increases as similarly discussed in quasi-static case;

14. the foregoing two aspects related to size-dependency in fatigue can be well described by the proposed SEL approach due to the consideration of the cyclic FPZ. Thanks to this approach, a size-dependent fatigue initiation of graphene nanocomposites in the Kitagawa-Takahashi diagram and a non-linear fatigue fracture scaling can be well captured, showing the transition from energy-based fatigue fracture to stress-based fatigue fracture as the specimen size decreases or the normalized initial crack length decreases. This is also an explanation of a longer fatigue lifetime to failure for a smaller specimen size as reported in this work for different graphene contents;

15. all the foregoing evidence showed that particular care should be devoted to the understanding of the size scaling of both quasi-static and fatigue fracture behaviors of general nanocomposites. The progress on this topic is very important for improving the design and guaranteeing the reliable applications of these materials in various engineering fields. To succeed, size effect testing and SEL analysis on geometrically scaled specimens can be adopted to provide reliable fracture properties, but not for the use of LEFM on a single specimen if the size is not large enough. A sufficiently large specimen to guarantee the use of the LEFM requires the size, depending on the materials, typically cannot be easily handled in the laboratory.

**Appendix 1. Compliance Method for Equivalent Crack Length**

To study fatigue crack initiation and propagation in graphene nanocomposites, cyclic growth of crack length must be properly quantified and carefully considered in the analysis. This is very important due to two aspects. One is that the measured crack length by a conventional camera is generally shorter than the real crack length due to the difficulties in detecting the extremely small crack opening close to the crack tip. This aspect can affect the accuracy of the measurement of the crack growth rate and SIF amplitude. The other is the FPZ formed by many small cracks ahead of a large crack in particular for quasi-brittle materials [62,63], meaning that a single crack does not exist. To overcome these issues, a method by leveraging the evolution of the specimen compliance can be used to estimate the growth of an equivalent crack length ($a_{eq}$) during the cyclic loading. The following relationship between the compliance ($c$) and the equivalent crack length can be obtained either by using FEA or analytical approach based on the equivalent LEFM:

$$c = \frac{9L^2}{2tD^2E^*} \int_0^\alpha g(\alpha)d\alpha + \frac{L^3}{48IE^*} \tag{14}$$

where $I$ is the second moment of area, $E^*$ is the elastic modulus for plain strain condition, and other symbols have the same meaning as mentioned in previous sections. Since the experimental fracture behavior of the material was successfully characterized by the equivalent LEFM, the crack length in Eq. (14) must be an equivalent crack length including the FPZ length. After knowing the equivalent crack length ($a_{eq}$), one can calculate the equivalent SIF amplitude by using the following relationship:

$$\Delta K_{eq} = \sqrt{G(\sigma_{max}, \alpha_{eq})E^*} - \sqrt{G(\sigma_{min}, \alpha_{eq})E^*} \tag{15}$$

It is worth mentioning here that the compliance function (Eq. (14)), for a given ratio of the span length $L$ to the specimen width $D$, is the same for two-dimensional geometrically scaled specimens. This is an important aspect for the investigation of the pure size scaling effect on the fatigue fracture of the material without being disturbed by any other factors (*e.g.*, notch effect, etc.). For instance, it is not possible to always create a sharp crack for some quasi-brittle materials (*e.g.*, concrete, rock, fiber-reinforced polymer composites, etc.). Instead, a blunt notch with a certain radius of curvature was made in these quasi-brittle materials for the investigation of their fatigue fracture behavior [57-60]. Accordingly, the notch effect can be possibly coupled with the size effect on the fatigue crack propagation in these materials since the results based on the compliance functions, having difference between notch and crack, can be misinterpreted. Thanks to the creation of a sharp crack for pure thermoset polymers and their graphene-nanomodified composites, the experimental results for the size effect on the fatigue fracture behavior in this study are not affected by the notch.

**Appendix 2. Energetic-equivalence Framework**

In addition to the proposed SEL method for the analyses of fatigue properties of quasi-brittle materials as discussed in Section 4.2, one may also use the method based on the energetic-equivalence framework [118,119] for the size effect data. The detail about this method is given next.

For energetic-equivalence framework, it is considered that the energy dissipation associated with the growth of the macro crack during each load cycle is equal to the sum of the energy dissipation associated with the propagation of all the active nano-scale cracks inside the cyclic FPZ. This assumption can be formalized as follows:

$$U_{c,\infty} da/dN = \sum_{i=1}^{n_a} U_a da_i/dN \quad (16)$$

where $U_{c,\infty}$ is the critical energy dissipation per unit growth of the macro crack in an infinitely large specimen, $U_a$ is the critical energy dissipation associated with the breakage of one atomic bond, $a_i$ is the length of $i$th nano-scale crack, and $n_a$ is the number of active nano-scale cracks in the cyclic FPZ.

By considering a well-established transition state theory proposed by Kramers [120], the fatigue crack growth rate at the nano-scale can be obtained in the following:

$$da_i/dN = v_i e^{-Q_0/kT} \Delta K_{ai}^2 \quad (17)$$

where $v_i$ is the constant determined by the geometry of the nano-scale element, $Q_0$ is the dominant free activation energy barrier, $k$ is the Boltzmann constant, $T$ is the absolute temperature, and $\Delta K_{ai}$ is the SIF amplitude of the nano-scale element. With the assumption on the average energy dissipations associated with the nano-scale cracks in the cyclic FPZ [118,119], the fatigue growth rate of a macro crack in a finite specimen can be written in the following form:

$$da/dN = \frac{A \Delta K^2}{U_c} \phi(\Delta K^2/(E U_c)) \quad (18)$$

where $A = \omega^2 U_a v_a e^{-Q_0/kT}$, $\omega = \Delta K_a/\Delta K$ represents the parameter connecting the nano-scale and macro-scale SIF amplitude, $\Delta K_a$ is the average SIF amplitude of the nano cracks in the cyclic FPZ, $\Delta K$ is the SIF amplitude of a macro crack, and $U_c$ is the critical energy dissipation per unit growth of the macro crack in a finite specimen.

By borrowing the idea on the SEL of quasi-static loading to the cyclic loading (i.e., $U_c = U_{c,\infty}[D/(D + D_{0,c})]$) where $D_{0,c}$ is the transitional size for the cyclic loading, the following equation can be obtained:

$$da/dN = \frac{A\Delta K_D^2}{U_{c,\infty}} \phi(\Delta K_D^2/(EU_{c,\infty})) = C\Delta K_D^m \quad (19)$$

where the function $\phi(\Delta K_D^2/(EU_{c,\infty})) = \Delta K_D^{2q}/E^q U_{c,\infty}^q$ assumes the self-similarity of crack propagation [121], $C = AE^{1-m/2}/U_{c,\infty}^{-m/2}$, $m = 2 + 2q$, and $\Delta K_D = (1 + D_{0,c}/D)^{0.5}\Delta K$ represents the adjusted SIF amplitude.

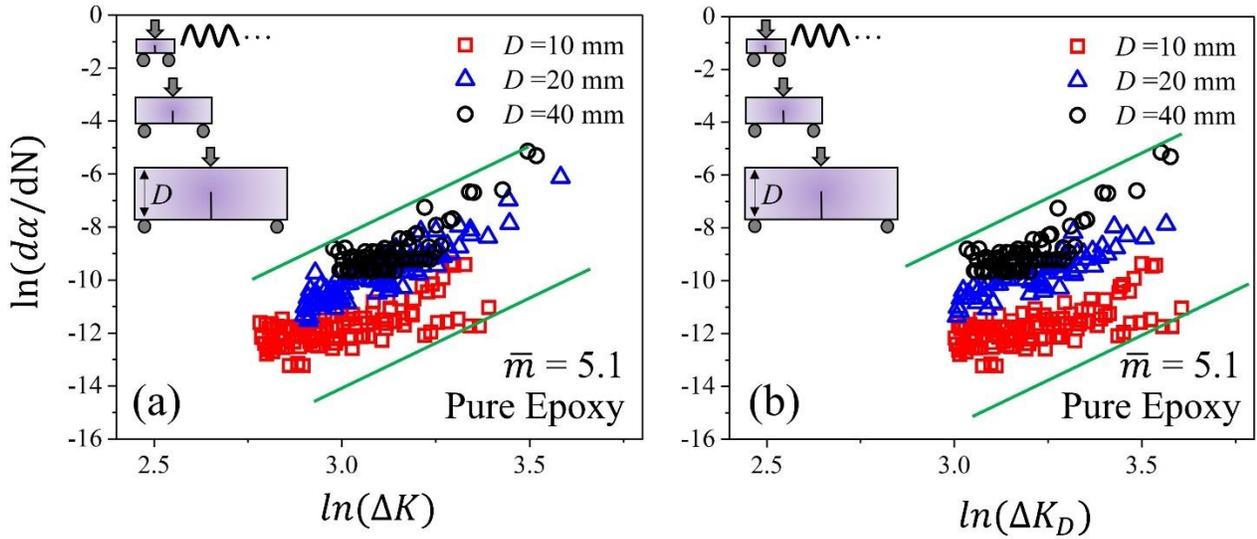

**Figure 18.** (a) Experimental Paris-Erdogan curves for pure epoxy with different sizes; (b) Adjusted Paris-Erdogan curves by leveraging Eq. (19) for pure epoxy with different sizes. Note that the cyclic growth of the crack length was directly measured by using a conventional digital microscope which does not really include the FPZ.

Now, let us plot the Paris-Erdogan curves of pure epoxy with different sizes as an example by using the LEFM-based SIF amplitude ($\Delta K = (1 - R)K = \sqrt{G(\alpha)E^*}$). In other words, the macro crack length during the cyclic loading in $\Delta K$ was directly measured via a conventional digital microscope which does not really include the FPZ. As illustrated in Figure 18a, the fatigue threshold and the parameter $C$ (i.e., intersection of the curve with the vertical coordinate in a double logarithmic scale) do not lead to the experimental data collapsed on a unique curve. It is worth mentioning here that the experimental data for the fatigue crack initiation stage was not plotted in Figure 18 since the small-scale cracks at that stage cannot be easily observed via a conventional detecting method. The foregoing phenomenon indicates that, while the LEFM can be used for the range of the investigated sizes under quasi-static loading for a good estimation, fatigue fracture behavior cannot be simply characterized by the LEFM due to enlarged cyclic FPZ.

By implementing Eq. ( 19 ) into the experimental data of pure epoxy as an example, the adjusted Paris-Erdogan curves with $\Delta K_D$ for pure epoxy with different sizes were plotted in Figure 18b. It is clear from this figure that the size effect on the fatigue threshold ($\Delta K_{th}$) and the critical SIF amplitude ($\Delta K_c$) can be removed. This adjustment requires the cyclic FPZ size near the fatigue threshold $c_{f,c} \approx 1$ mm, which is approximately 3 times larger than the monotonic FPZ size as mentioned in Section 4.1.2 and consistent with the estimation of the $c_{f,c}$ for pure epoxy by using Eq. ( 11 ) (*i.e.*, the proposed SEL for fatigue case). However, the adjusted Paris-Erdogan curves for pure epoxy still exhibit size dependence showing the noticeable differences on the parameter C. This aspect was not due to the frequency effect which requires significant difference on the applied frequencies as reported in the literature [122-125]. The main reason is related to the different stable evolution of the cyclic FPZ sizes for different specimen sizes in fatigue condition. This aspect is not considered in the adjusted Paris-Erdogan law by Eq. ( 19 ), which assumes a constant cyclic FPZ size.